\documentstyle[12pt]{dubna98}
\def\eqn{\begin{equation}}
\def\endeqn{\end{equation}}
\def\eqna{\begin{eqnarray}}
\def\endeqna{\end{eqnarray}}

\def\logmg{\ln\biggl({m\over\gamma}\biggr)}

\begin{document}

\vspace*{2cm}

\begin{center}
{\bf
NRQED APPROACH TO THE HYPERFINE STRUCTURE OF THE MUONIUM GROUND STATE
\footnote{Talk presented at the International Workshop $``$Hadronic Atoms
and Positronium in the Standard Model", Dubna, 26-31 May 1998}
\\}
\vspace*{1cm}
 T. KINOSHITA
\\ 
{\it  Newman Laboratory, Cornell University  \\ }
{\it  Ithaca, New York, 14853-5001  U.S.A. }
\end{center}


\vspace*{0.5cm}
\begin{abstract}
{\small
The method of NRQED 
is an adaptation of QED to bound systems.
While it is fully equivalent to QED,
it enables us to explore 
the QED bound states systematically
as an expansion in the fine structure constant $\alpha$
and velocity ($\sim Z\alpha$) of the bound electron.
I describe how to construct the
NRQED Hamiltonian choosing recent works on the 
$\alpha (Z\alpha )$, 
$\alpha^2 (Z\alpha )$, 
$\alpha (Z\alpha )^2$,
$\alpha (Z\alpha )^3$,
and $\alpha^2 (Z\alpha )^2$ radiative corrections
to the hyperfine structure
of the muonium ground state as examples.
}
\end{abstract}


\section{Introduction }
\label{sec:INTRO}

The most widely used approach to the relativistic
bound state problem is that based on the Bethe-Salpeter equation.
However, this tends to become very cumbersome
when applied to higher-order radiative corrections,
while the result reflects mostly the non-relativistic features
of the system, which are well described by the
Schr\"{o}dinger equation.
The problem is how to relate the latter
to the relativistic quantum field theory.
In the case of the electromagnetic interaction, 
a solution was given by the theory called non-relativistic 
quantum electrodynamics, or NRQED\cite{caswell}.

In this talk, I should like to describe the main features of NRQED
choosing as an example our recent work on the
radiative corrections to the hyperfine structure
of the muonium ground state. 
This quantity, which is
one of the most precisely measured quantities, is 
particularly suitable for a detailed 
examination of NRQED, being relatively free from 
the effects of hadronic interaction.  
The latest measurement gives
\cite{liu} 
\begin{equation}
  \Delta \nu (\mbox{exp})~ =~ 4~463~302.776~(51)~\mbox{kHz}~~~~~~
(11~\mbox{ppb}) ,        
\label{meas}
\end{equation}
which includes the earlier result \cite{mariam} and
reduces its uncertainty by a factor of three.


\section{The NRQED method}
\label{nrqed}

The NRQED is
a rigorous and systematic adaptation of QED to bound states.
It is described by a
Hamiltonian derived from QED
by expanding it in the velocity of the bound electron.
The equivalence of NRQED and QED  
is guaranteed by requiring that
the NRQED $scattering$ amplitude 
coincides with the QED $scattering$ amplitude
at some chosen momentum scale,
e.g., at the threshold of the external on-shell particles.
The coupling constant $\alpha$  and
mass,  renormalized on-shell, 
are identified with the observed values and used
as the input coupling constant and mass of NRQED.

The NRQED Hamiltonian consists of  all  possible local
interactions satisfying the required symmetries, such as
Galileian invariance, gauge invariance, parity invariance, time reversal invariance,
hermiticity,
and locality.
In the absence of dynamical photons it is nothing but
the Foldy-Wouthuysen-Tani (FWT) transform of the Dirac Hamiltonian.
Since the photon is always relativistic, 
its Lagrangian
is the same as that of QED.
When the interaction with photons is taken into account,
the NRQED Hamiltonian breaks up into two parts:
$H_{\rm main}^{\Lambda}$ and $H_{\rm contact}^{\Lambda}$.
$H_{\rm main}^{\Lambda}$
 is bilinear in fermion operators as well as photon operators. 
In addition, new photon interaction terms,
such as vacuum polarization
and light-by-light scattering,  
 are introduced to 
represent insertion of the fermion loop. 
$H_{\rm contact}^{\Lambda}$
represents the remainder.

The crucial feature of
NRQED is that all operators
(terms of the Hamiltonian) are restricted to
small momentum transfer, defined by some cut-off $\Lambda$,
while effects of large momentum transfer are represented
by coefficients of these operators
and terms of $H_{\rm contact}^{\Lambda}$.
Since this theory is meant to apply to non-relativistic systems,
the cut-off $\Lambda$ may be chosen as typical mass
scale of the system, e.g.,  the rest mass of the electron.

For the calculation of non-recoil terms in which
we can take the limit $m/M \rightarrow 0$,
$H_{\rm main}^{\Lambda}$ 
relevant to the muonium calculation 
can be written,
to order $\alpha$, as
\begin{eqnarray}
H_{\rm main}^{\Lambda}&=&\psi^{\dagger}(\vec{p}\,')
    \biggl [ { {\vec{p}\,^2} \over 2m } + eA^0
           - { (\vec{p}\,^2)^2 \over 8 m^3 }
           -{e \over 2m}  (\vec{p}\,'+\vec{p})\cdot\vec{A}
\nonumber \\
    &+&{e^2 \over 2m}  \vec{A}\cdot\vec{A}
    - {ie \over 2m} c_F\vec{\sigma}
      \cdot (\vec{q}\times\vec{A})
           - {e \over 8m^2} c_D\vec{q}\,^2 A^0
   +{ i e \over 4m^2} c_S\vec{\sigma}\cdot
                   (\vec{p}\,'\times\vec{p})A^0
\nonumber \\
    &-& { i e^2 \over 4m^2} c_S \vec{\sigma}\cdot
             (\vec{q_1}\times \vec{A}(q_1))  A^0(q_2)
   +{ i e \over 8m^2} c_S q^0 \vec{\sigma}\cdot
                   ((\vec{p}\,'+\vec{p})\times\vec{A})
\nonumber \\
 &+&{ie \over 8m^3} c_W(\vec{p}\,'^2+\vec{p}\,^2)\vec{\sigma}
                   \cdot (\vec{q}\times\vec{A})
     + {ie \over 8m^3} c_{q^2}\vec{q}\,^2\vec{\sigma}
                   \cdot (\vec{q}\times\vec{A})
\nonumber \\
    &+&{ie \over 8m^3} c_{p'p}
               \{ \vec{p}\cdot(\vec{q}\times\vec{A})
               (\vec{\sigma}\cdot\vec{p}\,')
                + \vec{p}\,'\cdot(\vec{q}\times\vec{A})
                     (\vec{\sigma}\cdot\vec{p}) \}
     +  \ldots~~  \biggr ] \psi(\vec{p})
\nonumber \\
  &+&c_{\rm vp}A^i(q){ \vec{q}\,^4 \over m^2 }
       A^j(q)(\delta^{ij}-{q^iq^j \over \vec{q}\,^2})
  +c_{\rm vp}A^0(\vec{q}){ -\vec{q}\,^4 \over m^2 }
     A^0(\vec{q})~,
\label{H_main}
\end{eqnarray}
where $\psi$ is the Pauli two-component
spinor,
$\vec{p}\,'$ and $\vec{p}$ are  the outgoing and incoming
fermion momenta, respectively, and $q=(q^0,\vec{q})$ is the 
incoming photon momentum. 
In the seagull vertex (on the third line), $\vec{q_1}$ is
the incoming momentum of the vector potential $\vec{A}$.
The large momentum-transfer effect is represented by the
NRQED $``$renormalization" coefficients which,
to order $\alpha$, are found to be
\begin{eqnarray}
c_F& =&1+ a_e~,
\nonumber \\
c_D&=&1 + { \alpha \over \pi } {8 \over 3}
     \biggl [ \ln \left({m \over 2\Lambda }\right)
-{3 \over 8}+{5 \over 6}\biggr ]
                 + 2 a_e~,
\nonumber \\
c_S&=& 1+ 2a_e ~,
\nonumber \\
c_W&=& 1~,
\nonumber \\
c_{q^2}&=& {\alpha\over\pi} {4 \over 3}\biggl [
\ln\left({m \over 2\Lambda}\right)
- { 3 \over 8 } + { 5 \over 6}  
+{1 \over 4}\biggr ]  + {a_e \over 2} ~,
\nonumber \\
c_{p'p}&=& a_e~,
\nonumber \\
c_{\rm vp}&=& {\alpha \over 15\pi } ~.
\label{Rconsts}
\end{eqnarray}
Note that $c_i$'s do not have coefficients 
involving $Z\alpha$
caused by the binding effect
because they are determined solely by comparison 
of the NRQED and
QED {\it scattering} amplitudes without  
referring to bound states.

$H_{\rm main}^{\Lambda}$ alone cannot reproduce 
the amplitudes of QED. 
To make NRQED fully equivalent to QED,  
we must add 
terms of contact interaction type:
\begin{eqnarray}
H_{\rm contact}^{\Lambda}= & - & d_1 {1 \over mM } (\psi^{\dagger} \vec{\sigma }\psi)
 \cdot (\chi^{\dagger} \vec{\sigma }\chi)
-d_2 {1 \over mM } (\psi^{\dagger} \psi) (\chi^{\dagger} \chi)
\nonumber \\
& - & d_3 {1 \over mM } (\psi^{\dagger} \vec{\sigma }\chi)
\cdot (\chi^{\dagger} \vec{\sigma }\psi)
 -d_4 {1 \over mM } (\psi^{\dagger} \chi )(\chi^{\dagger} \psi)
\nonumber \\
& - & d_5 {1 \over  m^3M } (\psi^{\dagger} 
\vec{D}^2\vec{\sigma }\psi)
\cdot  (\chi^{\dagger} \vec{\sigma }\chi) 
-~\cdots~~   ,   \label{contact}
\end{eqnarray}
where $\chi$ represents the muon (of mass $M$).
The coefficients $d_i$ are determined completely
by the requirement that 
the contact terms
make up the  difference between the  QED electron-muon  
scattering amplitude and the corresponding NRQED  scattering
amplitude generated by $H_{\rm main}^\Lambda$ 
and  lower order $H_{\rm contact}^\Lambda$.
Several terms of $H_{\rm contact}^{\Lambda}$ 
have been determined:
\begin{eqnarray}
d_1 &=&  \alpha(Z\alpha)^2 \pi {2\over 3} \left( \ln 2 -{13\over 4}
+{3\over 4} \right )
+ \alpha^2(Z\alpha)^2 {2\over 3} ( 0.767~9(79) ) 
+\cdots ~,
\nonumber \\
d_2 &=&  \alpha(Z\alpha)^2 4 \pi \left( { 139 \over 128 } + {5 \over 192} 
- {1\over 2} \ln 2 \right ) + \cdots ~,
\nonumber \\
 d_3 &=& d_4 = 0.
\label{hcontact}
\end{eqnarray}
The $Z\alpha$ factor in (\ref{hcontact}) 
comes from exchange of photons between 
the electron and the muon (of charge $-Ze$) and
has nothing to do with the bound-state effect.

Once the NRQED Hamiltonian is constructed
 the corrections to the energy 
and wave function can be determined
by the usual bound state
perturbation theory.


\section{Outline of non-recoil term evaluation}
\label{outline}

Thus far, non-recoil terms of order
$\alpha (Z\alpha )$, $\alpha (Z\alpha )^2$, and 
$\alpha^2 (Z\alpha )$ have been evaluated explicitly in NRQED.
They agree with evaluation by other methods.
Evaluation of $\alpha (Z\alpha )^3$ 
and $\alpha^2 (Z\alpha )^2$ terms
within the framework of NRQED is in progress.
Instead of discussing these calculations in detail,
I shall focus on their qualitative features
to illustrate
how the NRQED method works.
For simplicity let us discuss here only
diagrams with radiative photons,
leaving out the vacuum-polarization contribution.

In general, radiative corrections of order
$\alpha^m (Z\alpha )^n$ to the hyperfine splitting
are of two types:
$(A)$ 
First-order perturbation of contact terms 
of order $\alpha^m (Z\alpha )^{n+1}$.  $(B)$  
Higher-order perturbation of
lower-order terms of $H_{\rm main}^{\Lambda}$ and $H_{\rm contact}^{\Lambda}$.
The type $(A)$ originates from a set of QED scattering diagrams
in which $n+1$ photons are exchanged between the electron 
and the muon and the electron line is dressed by
$m$ virtual photon lines.

Since operators of the NRQED Hamiltonian describe
soft (low momentum-transfer) effects only,
the type $(B)$ scattering amplitude 
will reproduce the low energy behavior 
(including renormalization effects)
but not the complete QED scattering amplitude.
The difference between the QED scattering amplitude and the
NRQED scattering amplitude 
is due to large momentum-transfer
effect and leads to the type $(A)$ contact term.


{\bf The $\alpha (Z\alpha )$ term.}
The Hamiltonian $H_{\rm main}^{\Lambda}$ consists only of terms with
even parity.
Their expectation values  
with respect to the Coulomb wave function are even in $Z\alpha$, too.
Thus type $(B)$ terms cannot generate terms proportional to $Z\alpha$.
The only contribution is from the type $(A)$ term,
which arises from the difference
of QED and NRQED scattering amplitudes: 
\begin{equation}
-d_1 {1 \over {mM}} (\psi^\dagger \vec{\sigma}_e \psi )
  (\chi^\dagger \vec{\sigma}_\mu \chi ) \equiv
i {\cal T}^{QED} -
i {\cal T}^{NRQED} .
  \label{def-d1}
\end{equation}
The QED amplitude ${\cal T}^{QED}$ is derived from
the electron-muon scattering diagrams in which two photons
are exchanged between the electron and the muon 
and the electron line is dressed by a virtual photon line.
Both $ {\cal T}^{QED}$ and
${\cal T}^{NRQED}$
have threshold singularities
in the limit of vanishing
Coulomb photon momentum, which cancel in the 
difference (\ref{def-d1}).

The origin of the threshold singularity 
may be understood as follows:
In order to contribute to the hyperfine splitting, one of the two 
exchanged photons in the bound state perturbation theory
must be Coulomb-like while the other
is transverse.
The Coulomb photon can be absorbed in the wave function or the Green function.
Such a diagram is reduced to lower order in $Z\alpha$,
or, equivalently, multiplied by $1/(Z\alpha )$.
In the corresponding scattering state, this factor 
manifests itself
as an IR-divergent factor.
Since the bound state has no threshold singularity 
 all these singularities must cancel out in the end.

The expectation value of the contact term (\ref{def-d1}) can be evaluated 
in the first order bound state perturbation theory.
Only the value of the Coulomb wave function at the origin
contributes to the integral, leading to
\begin{equation}
\Delta \nu  = | \phi (0) |^2 {-d_1 \over {mM}} (\vec{\sigma_e} \vec{\sigma_\mu} )|_{J=0}^{J=1} ,
  \label{val-d1}
\end{equation}
where $| \phi (0) |^2 = \gamma^3 /\pi$ for the ground state.
The difference between $J = 1$ and $J = 0$ spin states is taken.
Together with the vacuum-polarization contribution
this gives the first term of $d_1$ in (\ref{hcontact}).


{\bf The $\alpha^2 (Z\alpha )$ term.}
For the same reason as in the $\alpha (Z\alpha )$ case
this correction comes only from the contact term of type $(A)$.
It is defined as the difference between the gauge invariant set
of QED scattering diagrams in which the electron and the muon 
exchange two photons
and the electron line is dressed by two radiative photons
and the corresponding NRQED scattering diagrams.


{\bf The $\alpha (Z\alpha )^2$ term.}
One contribution comes from a type $(A)$ contact term of order 
$\alpha (Z\alpha )^3$, which arises from
QED scattering diagrams 
that exchange three photons between the electron and the muon,
dressed by one radiative photon on the electron line.
These amplitudes have threshold singularity of
up to order $(m/\lambda )^2$.
The type $(B)$ NRQED scattering diagrams which have the same leading threshold
singularity are of two kinds: 
One is the first term of $d_1$
in (\ref{hcontact}) combined with one Coulomb potential
and the other is the Fermi potential times the anomalous
magnetic moment combined with two Coulomb potentials.


{\bf The $\alpha^2 (Z\alpha)^2$ term.}
The relevant QED diagrams are those that exchange three photons
between the electron and the muon, dressed by two
radiative photons on the electron line.
The corresponding NRQED diagrams have two radiative 
photons, too.
However, these diagrams contribute only to order 
$\alpha^2 (Z\alpha )^4$, which is too small by a factor
$(Z\alpha )^2$.
Thus the radiative correction of order $\alpha^2 (Z\alpha )^2$
comes from the QED diagrams only.
The contact term of NRQED derived from these QED diagrams
does not have the $\ln (m/\gamma )$ dependence.
Thus logarithmic terms of order $\alpha^2 (Z\alpha )^2$
can come only from the NRQED operators of $H_{\rm main}^{\Lambda}$
contributing to the $\alpha (Z\alpha )^2$ term
combined with an additional factor of $\alpha$
found in the $``$renormalization" effect.
This leads to the result reported in 
(\ref{a^2(Za)^2}).


{\bf The $\alpha (Z\alpha)^3$ term.}
The status of the theory of this term will be
discussed in detail by Dr. Nio in her talk presented 
at this Workshop.


\section{Current status of theory of muonium hfs}
\label{sec:muhfs}

As is well known, the bulk of the hyperfine splitting 
is given by the Fermi formula  
\begin{eqnarray}
  E_F&=&{16 \over 3}\alpha^2 c R_{\infty} {m \over M}
  \left [ 1+{m \over M} \right ]^{-3}   \nonumber  \\
     &=& 4~453~839.405~(518)~{\rm kHz},    
       \label{EF}
\end{eqnarray}
where $R_{\infty}$ is the Rydberg constant 
for infinite nuclear mass,
and $m$ and $M$ are the electron 
and muon masses, respectively.  
The values used above for $\alpha$,
$R_{\infty}$, and $M/m$ are \cite{hk,udem,liu}
\begin{eqnarray}
  \alpha^{-1}&=&137.035~999~58~(52)~~~(3.8~\rm{ppb}),    \nonumber   \\
  R_{\infty} &=&10 ~973~~731.568~639~(91)~ {\rm m}^{-1},  \nonumber   \\
  {M  \over  m } &=& 206.768~273~(24) ~~~~~(117~\rm{ppb}) .
 \label{constants}
\end{eqnarray}

Various theoretical corrections to $E_F$ have
been calculated over the last 50 years.
They may be classified into non-recoil and recoil terms,
both including radiative and binding effects.
In addition there are contributions of hadronic 
vacuum-polarization and weak interaction effect.  
Thus one may write
\begin{equation}
  \Delta \nu (\mbox{theory}) = \Delta \nu (\mbox{non-recoil}) 
  + \Delta \nu (\mbox{recoil}) 
 + \Delta \nu (\mbox{hadron}) 
  + \Delta \nu (\mbox{weak}) .    \label{theoryform}
\end{equation}

A purely radiative non-recoil term of order $\alpha (Z\alpha )$ 
and an approximate value of the $\alpha (Z \alpha)^2$ term have been known 
for some time \cite{SY}.
Recently radiative corrections of orders $\alpha (Z\alpha)^2$
and $\alpha^2 (Z\alpha )$ have been evaluated with high precision.
Including these results
and the Fermi term $E_F$ itself, the non-recoil 
term of the muonium hyperfine structure 
can be written as  
\cite{nk1,kn1,es1,kn2,pach1,karshenboim}
\begin{eqnarray}
  \Delta \nu (\mbox{non-recoil}) &=& E_F (1 + a_{\mu} )
  \left (  1 + b + a_e + \alpha (Z \alpha )
\left ( \ln 2 - {5 \over 2} \right ) \right .   
\nonumber   \\ 
&-&  \left . {{8 \alpha (Z \alpha )^2} \over {3 \pi}} 
\ln (Z\alpha) \left [ \ln (Z\alpha) - \ln 4 + {281 \over 480} 
\right ] \right .    \nonumber  \\
&+&  \left . 16.9042 (11) {{\alpha (Z\alpha )^2} \over {\pi}} 
             +  0.7717 (4) {{\alpha^2 (Z\alpha )} \over {\pi}} 
 \right )   \nonumber   \\
     &=& 4~464~098.677~(518)~{\rm kHz}.
\label{nonrecoil}
\end{eqnarray}
Here $a_e$ and $a_{\mu}$ are the anomalous magnetic 
moments of the electron and muon, respectively, and $b$ is the Breit term  
\begin{equation}
  b = {1 \over {\beta (2\beta -1)}} - 1,~~~\beta = (1 - (Z\alpha )^2 )^{1/2}.
 \label{breit-term}
\end{equation}
The appearance of the factor $(1 + a_{\mu} )$ in 
(\ref{nonrecoil}) is in accord with the definition 
(\ref{EF}) of $E_F$.
The coefficient 16.9042 of the $\alpha (Z\alpha )^2$ 
term is different from 15.39 given in \cite{SY}.
This is because the latter contains, besides 
the $\alpha (Z\alpha )^2$ term,
terms of order
$\alpha (Z\alpha )^3$ and higher. 
More recent evaluation of this term without expansion
in $Z\alpha$ is given in \cite{blundell}.
For a detailed discussion see  \cite{kn2}.

In order to match the precision
of the new measurement \cite{liu}, however,
it is necessary to improve the theoretical prediction
of terms of order $\alpha^4$.
This is because,
when enhanced by the factor 
$\ln^2 (m/\gamma )$ or 
$\ln(m/\gamma )$ where $m/\gamma  = 1/Z\alpha $, 
corrections of order
$\alpha (Z\alpha )^3$ and $\alpha^2 (Z\alpha )^2$
contribute to the muonium 
hyperfine structure as much as the $\alpha^2 (Z\alpha )$ 
term.
(There is no $\ln (Z\alpha )$ enhancement for $\alpha^4$, 
$\alpha^3 (Z\alpha )$ and $(Z\alpha )^4$ terms.)

The leading log contribution of the $\alpha (Z\alpha )^3$ term is 
known exactly \cite{kn1,lepage,karsh2,nk2}.
The non-log term has also been estimated \cite{nk2}.
Altogether we have
\eqna
\Delta \nu (\alpha(Z\alpha)^3)
&=&  \alpha (Z\alpha)^3  E_F \logmg 
 \biggl [  
       \left ( 5\ln 2-{ 191 \over 16 }  \right)
        +    {13 \over 24}  \biggr ] ~
\nonumber \\
&-&  (12 \pm 2 ) \alpha (Z\alpha)^3  E_F /\pi
\nonumber \\
  &=& -0.542~(8)~\mbox{kHz} ,  
\label{a(Za)^3log}
\endeqna
where the first term in the brackets is the 
contribution from the radiative photon 
and the second is due to 
the vacuum polarization.
The uncertainty is an estimate based on the non-log part of
the $\alpha (Z\alpha )^3$ term given in \cite{blundell}.
We are trying to improve this uncertainty.

For the $\alpha^2 (Z\alpha )^2$ term
the leading log part together with an estimate of non-log part 
gives the value \cite{karshenboim,nk2}
\eqna
\Delta \nu (\alpha^2(Z\alpha)^2)&=&{\alpha^2(Z\alpha)^2 \over \pi^2}  E_F 
\ln^2 \left ( {m \over \gamma } \right )
\biggl(-  {4 \over 3}\biggr ) 
\biggl(1-{4\pi m \over \alpha M }\biggr)  
\nonumber   \\ 
&+&{\alpha^2(Z\alpha)^2 \over \pi^2}  
 E_F \logmg 
 \biggl [ -2 \biggl( -{4358\over1296}
-{10\over27}\pi^2 +{3\over2}\pi^2 \ln2 -{9\over4}\zeta(3) \biggr) 
\biggr .
\nonumber \\ 
\biggl . &-& {3\over2}\biggl( {197\over144}+\biggl({1\over2} - 3\ln2\biggr)\zeta(2)
+ {3\over4}\zeta(3) \biggr)
\nonumber \\ 
&+& \biggl(1-{4\pi m \over \alpha M }\biggr)
         {8\over3}\biggl(-\ln2+{3\over4}\biggr)
+{1\over2}\biggl( -{11\over9} -1+{8\over15} \biggr) \biggr]
\nonumber \\  
&+&{\alpha^2(Z\alpha)^2 \over \pi^2} E_F
\biggl(1-{4\pi m \over \alpha M }\biggr) (10 \pm 2.5) 
\nonumber \\  
&=&0.193~(24)~\mbox{kHz} ,   
\label{a^2(Za)^2}
\endeqna
where the factor $m/M$ originates from the reduced mass effect.
An error in \cite{kn1} due to an incorrect treatment
of low energy contribution is rectified in (\ref{a^2(Za)^2}).

The pure recoil and order $\alpha$ radiative recoil corrections,
except for the last term of order $(m /M )^2$
obtained recently \cite{eides1},
 have been known for some time
\cite{SY,EKS0,EKS1} 
\begin{eqnarray}
 \Delta \nu (\mbox{recoil})&=&E_F \left ( 
- {{3Z\alpha} \over {\pi}} {{m M} \over 
{M^2 -m^2 }} \ln {M \over m} 
    + {{\gamma^2} \over {m M}} 
\left [ 2 \ln {{m_r} \over {2\gamma}} - 6 \ln 2 
+ {65 \over 18} \right ] \right )   
\nonumber   \\
&+& E_F {{\alpha (Z\alpha)} 
\over \pi^2} {m \over M} \left ( - 2 \ln^2 
{M \over m} + {13 \over 12} \ln {M \over m}
    \right .   
\nonumber   \\
&+&\left .  6 \zeta (3) + \zeta (2) - {71 \over 72} 
+ 3 \pi^2 \ln 2  
 + Z^2 \left [ {9 \over 2} \zeta (3) 
+ {39 \over 8}- 3\pi^2 \ln 2 \right ]   \right .  
\nonumber   \\ 
&+&  \left .  {\alpha \over \pi} \left [ - {4 \over 3} 
\ln^3 {M \over m} + {4 \over 3} 
\ln^2 {M \over m}  
+ {\cal O} \left ( \ln {M \over m} \right ) 
\right ] \right )      
\nonumber   \\
&+&   E_F \alpha (Z\alpha ) \left ( {m \over M } \right )^2 
           \left [ \left (-6 \ln 2 - {3 \over 4}\right ) 
                   - {17 \over 12} Z^2  \right ]
\nonumber   \\ 
  &=& - 795.228~(2)~{\rm kHz},     
           \label{recoil}
\end{eqnarray}
where $\gamma \equiv Z \alpha m_r ,~ m_r 
=  m M /(m + M )$.
The $\ln^3$ and $\ln^2$ parts of the $\alpha^2 (Z\alpha )$ 
term were evaluated by Eides $et~al$. \cite{EKS1}.    

Further refinement of $\Delta ({\rm recoil})$
comes from evaluation of $(Z\alpha )^3 (m/M)$ terms.
One of the leading log term of this contribution arises from
the recoil hfs potential 
(the first line of (\ref{recoil})) and the
1-photon exchange part of the Coulomb Green function:
\eqna
\Delta \nu^{(1)} ((Z\alpha)^3(m/M))&=& 2 < V_{\rm hfs} G_{1p} V_K > +2 < V_{\rm hfs} G_{1p} V_D >
\nonumber \\
 &\simeq&   - { 3 (Z\alpha)^3 \over  \pi} { m \over M}
   E_F \ln \left({M \over m}\right) \logmg~  
\nonumber   \\  
 &=&  -0.210~(43)~\mbox{kHz} ,  
\label{hfsKD} 
\endeqna
where $V_K$ and $V_D$ are the $k^4$ kinetic energy and Darwin terms,
respectively.
The uncertainty is estimated assuming
that it is a factor $\ln(Z\alpha)$
smaller than the leading term.

Another log term arises from the Salpeter 
recoil correction to the Lamb shift \cite{salpeter}.
This contribution involves both relativistic and 
non-relativistic regions.
Unfortunately, the latter was handled incorrectly in \cite{kn1}.
Correcting this error we find \cite{karshenboim,nk2}
\eqna
\Delta \nu^{(2)} ((Z\alpha)^3(m/M))&=& 
{(Z\alpha)^3 \over \pi} E_F { m\over M}
 \ln^2 \left ( {m \over \gamma } \right ) 
\biggl[ - {2 \over 3} \biggr ]
\nonumber \\
&+& {(Z\alpha)^3 \over \pi} E_F { m\over M}
\logmg \biggl[ -2C_S   
+ {32\over3} \left(- \ln2 + {3\over4} \right) \biggr ] 
\nonumber \\
&+& {(Z\alpha)^3 \over \pi} E_F { m\over M} (40 \pm 22)
\nonumber \\
   &=& -0.194~(59)~\mbox{kHz} , 
\label{(Za)^3m/M_2}
\endeqna
where one factor of $Z\alpha$ is radiative correction
due to exchange of transverse photon between the electron
and the muon and $C_S$ is given by
\begin{equation}
C_S = -{1 \over 9} + {7 \over 3} (  2 \ln 2 + 3)   
-  {2 \over {1 - (m/M)^2}} 
\left ( \ln (1 + m/M ) - (m/M)^2 \ln (1 + M/m ) \right ) .
\label{C_Sdef}
\end{equation}

The hadronic vacuum polarization contributes \cite{KF}
\begin{eqnarray}
\Delta \nu (\mbox{hadron}) &=& {{\alpha (Z\alpha )} \over \pi^2 } 
{{ m M } \over m_\pi^2 } (3.75 \pm 0.24 ) E_F  \nonumber \\
    &=& 0.250~(16)~{\rm kHz} ,
\end{eqnarray}
where $m_\pi$ is the charged pion mass.

Finally, there is a contribution due to the $Z^0$ exchange  
\cite{BF,BF2,privatecom}
\begin{eqnarray}
  \Delta \nu (\mbox{weak}) 
&=& -G_F {{3 \sqrt{2} mM} \over {8 \alpha \pi}} E_F \nonumber \\
   &\simeq& -  0.065 ~\mbox{kHz},   \label{weak}
\end{eqnarray}
where $G_F$ is the Fermi weak coupling constant.

Collecting all contributions we obtain
the best current estimate
\begin{equation}
 \Delta \nu (\mbox{theory}) 
 = 4~463~302.649~(517)~(34)~(\leq 100)
  \mbox{kHz}  ~,    \label{newtheory1}
\end{equation}
where the first and second errors come from those of $M/m$
and $\alpha$ in (\ref{constants}) and the last
is an estimate of the theoretical uncertainty
from the yet-to-be calculated terms.
This is in excellent agreement with the measurement (\ref{meas}).


\section{Concluding remarks}
\label{sec:remarks}

Current uncertainties in (\ref{a(Za)^3log}), (\ref{a^2(Za)^2}), (\ref{hfsKD}), and
(\ref{(Za)^3m/M_2}) are 0.008 kHz, 0.024 kHz, 0.043 kHz, and 0.059 kHz,
respectively.
For further improvement it is particularly important to 
evaluate the 
$\cal{O}$($\alpha^3 (m/M)$) recoil terms.

Study of positronium in the framework of NRQED 
requires additional terms to the Hamiltonian.
This is discussed by P. Labelle in this Workshop.
Adaptation of NRQED method to the chiral perturbation theory
is of great interest at present in view of the fine
experimental works on the $\pi^+ \pi^-$ bound system.

This work is supported in part by the U. S. National Science Foundation.


\end{document}